\documentclass[11pt]{article}

\usepackage[square,numbers]{natbib}

\usepackage{amsmath, amssymb, color, epsf, epsfig, graphicx, subfigure}

\makeatletter

\usepackage[T1]{fontenc}
\usepackage[latin1]{inputenc}

\makeatother

\begin{document}

\setcounter{footnote}{4}

\title{\bfseries A theory of Plasma Membrane Calcium Pump stimulation and activity}

\author{Michael Graupner$^{1,2}$\footnote{To whom correspondence should be
    addressed, current address: Laboratoire de Neurophysique et
    Physiologie, Universit\'e Ren\'e Descartes - Paris V, 45, rue des
    Saints P\`eres, 75270 Paris Cedex 06, France, Tel.:
    +33-(0)1.42.86.38.13, E-mail: michael.graupner@univ-paris5.fr}$\;$,
  Frido Erler$^1$ and Michael Meyer-Hermann$^{1,3,4}$ \\[.2cm]
  {\it $^1$Institute for Theoretical Physics, Dresden University of Technology,} \\
  {\it 01062 Dresden, Germany}\\[1mm]
   {\it $^2$Laboratoire de Neurophysique et Physiologie, CNRS UMR
     8119,} \\
   {\it Université René Descartes - Paris V } \\
   {\it 45, rue des
    Saints P\`eres, 75270 Paris Cedex 06, France}  \\[1mm]
  {\it $^3$Centre for Mathematical Biology, Mathematical Institute, 24-29 St.~Giles,}\\
  {\it Oxford University, Oxford OX1 3LB, United Kingdom} \\[1mm]
  {\it $^4$Frankfurt Institute for Advanced Studies (FIAS), Johann Wolfgang }\\
  {\it Goethe-University, Max von Laue-Str.~1, 60438 Frankfurt/Main,
    Germany}
}


\date{Journal of Biological Physics 31:183-206, 2005}

\maketitle

\begin{abstract} 
\noindent
The ATP-driven Plasma Membrane Calcium pump or Ca$^{2+}$-ATPase (PMCA)
is characterized by a high affinity to calcium and a low transport
rate compared to other transmembrane calcium transport proteins. It
plays a crucial role for calcium extrusion from cells.  Calmodulin is
an intracellular calcium buffering protein which is capable in its
Ca$^{2+}$ liganded form of stimulating the PMCA by increasing both the
affinity to calcium and the maximum calcium transport rate. We
introduce a new model of this stimulation process and derive
analytical expressions for experimental observables in order to
determine the model parameters on the basis of specific experiments.
We furthermore develop a model for the pumping activity. The pumping
description resolves the seeming contradiction of the Ca$^{2+}$:ATP
stoichiometry of 1:1 during a translocation step and the observation
that the pump binds two calcium ions at the intracellular site.  The
combination of the calcium pumping and the stimulation model correctly
describes PMCA function.  We find that the processes of
calmodulin-calcium complex attachment to the pump and of stimulation
have to be separated. Other PMCA properties are discussed in the
framework of the model. The presented model can serve as a tool for
calcium dynamics simulations and provides the possibility to
characterize different pump isoforms by different type-specific
parameter sets.

\end{abstract}



\section*{Introduction} 

The Plasma Membrane Calcium pump with a high calcium affinity
($K_{1/2} < 0.5 - 1 \, \mu\rm M$ 
\cite{Zylinska2000, Penniston1998, Carafoli1991}) and a low transport
rate ($\approx 30 \, \rm Hz$, \cite{Juhaszova2000} and private
communication) is an important component for the maintenance of
calcium homeostasis in cells.  By using the energy stored in ATP
the PMCA transports intracellular calcium ions out of the cell. It has
been found in all mammalian cells \cite{Guerini1998}, where the
expression level does not exceed 0.1\% of the total membrane protein
\cite{Zylinska2000, Stauffer1995, Carafoli1992, Chin2000}. An
exception is the brain where this value is up to 10 times higher than
in non-excitable cells \cite{Guerini1998}.

The four different pump isoforms are encoded by four independent
genes, which are indicated by numbers 1-4. The diversity of pump forms
is further increased by alternative mRNA splicing variants,
characterized by small letters. There exist more than 26 transcripts
which differ in their regulatory properties, for instance in their
affinity to calmodulin, and which are distributed in a tissue specific
manner \cite{ Zylinska2000, Carafoli1991}. Referring to the available
data we will investigate hPMCA2b and hPMCA4b, where
h stands for human \cite{Caride2001b}.

Calmodulin is an intracellular calcium sensor protein with four
relatively high affinity Ca$^{2+}$ binding sites ($K_a = 1-10 \,\,
\mu$M$^{-1}$ at low ionic strength \cite{Linse1991}). It belongs to
the mobile proteins of the EF-hand family with a helix-loop-helix
conformation \cite{Pottorf2000}. Two calcium ions are bound at the
N-terminal as well as at the highly homologous C-terminal domain, each
of them formed by two EF-hands \cite{Chin2000}. The domains are
connected by a flexible linker - an $\alpha$-helix \cite{Chin2000}.
Because of its calcium binding capacity calmodulin becomes relevant
for the spatial propagation of calcium signals within the cytoplasm.
Additionally the fully liganded calmodulin-calcium complex is
responsible for Ca$^{2+}$-dependent regulation of the activities of a
vast array of different target proteins, including enzymes, ion pumps
and channels \cite{Persechini1999}. Among those with high affinity
($K_d = 5$ nM) to the calmodulin-calcium complex is the PMCA
\cite{Meyer1992}. The free calmodulin-binding domain of the PMCA also
interacts with the ATP-binding site of the pump and acts as an
inhibitor of ATP-driven pump function \cite{Carafoli1992,
  Persechini2000a}.  The detachment of that autoinhibitory domain
segment after binding of the complex causes a stimulation of the pump
function by increasing both the affinity to calcium and the maximum
turnover rate of calcium \cite{Penniston1998, Carafoli1991, Chin2000,
  Elwess1997}. The stimulation and the relaxation to the initial
unstimulated state happens on a time scale of minutes and enables the
pump to display a memory of previous calcium transients
\cite{Caride2001b}.

Caride {\it et al.} have published a stimulation model making use of
measurements of the dynamical stimulation behavior \cite{Caride2001b}.
We extend the reaction scheme and aim to improve the model results in
two ways: Firstly, we will deduce the stimulation parameters from
measured data.  Secondly, we will include a saturation of the
stimulation rate for high calcium concentrations.  Indeed, measured
stimulation constants call for a limitation by a maximum value at high
calcium concentrations. After completion of this work we were made
aware of the recently published article in which Penheiter {\it et
  al.} propose a new stimulation model in conjunction with
fluorescence measurements \cite{Penheiter2003}.  Although Penheiter's
model includes saturation, the relaxation of the PMCA to resting state
is not considered.  We separate the stimulation as well as the
relaxation into two steps and introduce rate limiting reactions in our
model.  An analytical approach enables us to deduce the required rate
constants from measurements.  The inferred system of coupled
nonlinear differential equations is solved
numerically. The presented new model is in agreement with the
investigated experimental data of the stimulation process by
calmodulin. Furthermore we predict the stimulation behavior beyond the
available data.

Besides the stimulation model we also consider the calcium pumping
activity. This joined model of stimulation and pumping may serve as a
tool for simulations in a wide variety of systems such as single cells
or tissues.  For that purpose we assume that all isoforms are
functionally similar and characterize the different isoforms by type
specific sets of parameters.  In this way we differentiate between
universal system independent parameters which are characteristic for
each pump type and specific parameters such as the PMCA protein
expression level, which has to be adapted to each experimental setup.
In relation to the available experiments we can specify parameter sets
for the h2b and h4b isoform which turn out to be in a biologically
reasonable range.

We begin by introducing the stimulation model reactions. We determine
the model parameters and calculate stimulation dynamics. In the second
part of this paper we focus on the derivation of an expression for the
calcium pumping activity. Finally the stimulation and the pumping
model together serve to fully describe PMCA function.


\section*{Model}

\subsection*{Stimulation model}

\setcounter{footnote}{1} We base the stimulation model on a system of
rate reactions, which describe the calcium and calmodulin dependent
transition from unstimulated to stimulated pump form and vice versa.
The stimulation dynamics of an ensemble of PMCA's is characterized
by the following reactions: \\
\parbox{.5\textwidth}{
\begin{eqnarray} 
 \rm M + 4 Ca^{2+}&\overset{k_1}{\underset{k_{-1}}{\rightleftharpoons}} & \rm X_4 \label{buff},  \\%
 \rm P + X_4 & \overset{k_2}{\underset{k_{-2}}{\rightleftharpoons}} &  \rm P_{\rm X} \label{bind}, \\%
 \rm  P_{\rm X} & \overset{k_3}{\rightharpoondown} &  \rm  P_{\rm X}^* \label{stim}, 
\end{eqnarray}}
\parbox[c]{.5\textwidth}{
\begin{eqnarray}
 \rm P_{\rm X}^* & \overset{k_4}{\underset{k_{-4}}{\rightleftharpoons}} & \rm P^* + X_4 \label{detach},\\%
  \rm P^* & \overset{k_5}{\rightharpoondown} & \rm P \label{unstim}.  
\end{eqnarray}}
We use the following abbreviations: P denotes free PMCA, $\rm CaM
\negmedspace \cdot \negmedspace Ca_4 \equiv X_4$, $\rm P \negmedspace
\cdot \negmedspace CaM \negmedspace \cdot \negmedspace Ca_4 \equiv
P_{\rm X}$, $ \rm P^* \negmedspace \cdot \negmedspace CaM \negmedspace
\cdot \negmedspace {Ca_4} \equiv P_{\rm X}^*$, and $\rm P^*$ free
stimulated pump. P and $\rm P_{\rm X}$ are the unstimulated pumps and
the asterisk denotes stimulated pumps. M and Ca$^{2+}$ are free
intracellular calmodulin and calcium, respectively.  Note that in this
article Roman style symbols refer to elements or proteins whereas
italic style symbols denote concentrations or fractions of the
respective element or protein. The $k$'s are the rate constants.  With
the Law of Mass Action the model \eqref{buff}-\eqref{unstim} is
rewritten as a set of differential equations for concentrations (see
equations \eqref{eqn:diff_p}-\eqref{eqn:diff_p*} in the Appendix for
the stimulation model).

Reaction \eqref{buff} describes the binding of calcium to calmodulin.
The cooperative binding of Ca$^{2+}$ within each domain of the
calmodulin protein \cite{Linse1991} is simplified in this reaction by
the assumption of highly cooperative binding of all four calcium ions.
Based on the realistic assumption that the binding of calcium occurs
faster than the PMCA stimulation, which happens on a time scale of
minutes, we treat this binding to be in
quasi-steady-state.

The assumption of irreversible stimulation \eqref{stim} and relaxation
\eqref{unstim} implies that the pump can only be stimulated when X$_4$
has attached to the pump, whereas recovery back to the unstimulated
pump form only occurs when X$_4$ has detached. It will be seen later
on that the irreversible step \eqref{stim} and the detachment step
\eqref{detach} limit the stimulation and relaxation rate, {\it i.e.}
both cannot exceed the rates $k_3$ and $k_4$, respectively. Note that
\eqref{detach} does not necessarily mean that the complex detaches as
a whole. In principle the calcium ions could also detach from P$_{\rm
  X}^*$ first without any alterations of the model because of the
quasi-steady-state approximation in \eqref{buff}.  Thus, the essential
statement of \eqref{detach} is that calmodulin (with or without
calcium) has to detach in order to destimulate the pump.

Note that we introduced different pathways for the stimulation
(reactions \eqref{bind} and \eqref{stim}) and the relaxation process
(reactions \eqref{detach} and \eqref{unstim}). This enables us to describe
the temporal behavior of these two processes independently of each other,
which is rather important in order to reflect isoform specific
stimulation and relaxation dynamics.


\paragraph{Stimulation model parameter determination}  

We aim to relate the rate constants $k_i$ \, $(i = \pm1, \pm2,3, \pm4,
5)$ and the stimulation and relaxation constants $k^{\rm exp}_{\rm
  stim}$ and $k^{\rm exp}_{\rm relax}$ measured in experiment by
Caride {\it et al.}  \cite{Caride2001b}. Note that in
\cite{Caride2001b} these constants are denoted by $k_{\rm act}$ and
$k_{\rm inact}$, respectively.  Starting from the stimulation model we
deduce analytical expressions for the exponential growth constants
$k_{\rm stim}$ and $k_{\rm relax}$.


\paragraph{Stimulation}

Suppose only unstimulated pump form P to be present with the
subsequent addition of calmodulin.  In the very beginning the
stimulation model can be reduced to the rate equations \eqref{buff},
\eqref{bind} and \eqref{stim}.  The binding of calcium to calmodulin
(reaction \eqref{buff}) is assumed to be in quasi-steady-state,
therefore $X_4 \equiv CaM \negmedspace \cdot \negmedspace Ca_4 = K
\cdot M \cdot {Ca}^4$, with $K=k_1/k_{-1}$.  During the beginning of
the stimulation the finite calmodulin concentration imposes no
restrictions on the dynamics, since only a minor fraction of free
X$_4$ is bound.  Therefore, the conservation of the calmodulin
concentration is neglected and $M$ is considered to be constant.  With
this assumption and the pump mass conservation $P_0 = P + P_X + P_X^*$
(note that $P^*$ is not involved here) reactions \eqref{bind} and
\eqref{stim} can be rewritten as a system of three linear homogeneous
differential equations of first order (the full nonlinear system is
shown in equations \eqref{eqn:diff_p}-\eqref{eqn:diff_p*} in the
Appendix).  The system is solved via the ansatz $P = A_1 \cdot
exp(-k_{\rm stim}t)$, $P_X = A_2 \cdot exp(-k_{\rm stim}t)$ and $P_X^*
= A_3 \cdot exp(-k_{\rm stim}t)$. The characteristic equation leads to
three solutions for $k_{\rm stim}$: $k_{\rm stim}^1 = 0$ and
\begin{equation}
k_{\rm stim}^{(2/3)} =  \frac{k_{-2} + k_3 + k_2 X_4}{2} \pm \sqrt{\left(\frac{k_{-2} + k_3 + k_2 X_4}{2}\right)^2 - k_2 k_3 X_4} \label{eqn:sol_k_stim}.
\end{equation}
With the boundary conditions $P(0) = P_0$, $P( \infty )=0$, $P_X(0)
= 0$, $P_X(\infty) = 0$, $P_X^*(0)=0$, $P_X^*(\infty) = P_0$ and the pump
mass conservation $P_0 = P(t) + P_X(t) + P_X^*(t)$  $\forall \, \, t$,
the exact solution for the stimulated pump form becomes
\begin{equation}
P_X^*(t) = P_0 \left[ 1 + \left( \frac{k_{\rm stim}^{(2)}}{k_{\rm stim}^{(3)}} - 1 \right)^{-1} e^{-k_{\rm stim}^{(2)} t} - \left( 1-\frac{k_{\rm stim}^{(3)}}{k_{\rm stim}^{(2)}} \right)^{-1} e^{-k_{\rm stim}^{(3)} t} \right].
\label{eqn:stim_stimform_exact}
\end{equation}

During the first phase of stimulation the pump exhibits a single
exponential behavior as discussed in \cite{Caride1999}. The exact
solution for $P_X^*(t)$ in equation \eqref{eqn:stim_stimform_exact} can
be simplified to a single exponential expression in two
cases. With $k_{\rm stim}^{(2)} >> k_{\rm stim}^{(3)}$ equation
\eqref{eqn:stim_stimform_exact} becomes
\begin{equation}
P_X^*(t) = P_0(1-e^{-k_{\rm stim}^{(3)} t}).
\label{eqn:stim_stimform_estimat_1}
\end{equation}
Formally, also $k_{\rm stim}^{(2)} << k_{\rm stim}^{(3)}$ yields a
single exponential behavior of the same form but with $k_{\rm
  stim}^{(2)}$ in the exponent. However, $k_{\rm stim}^{(2)} \ge
k_{\rm stim}^{(3)}$ holds true (see equation \eqref{eqn:sol_k_stim}).

Assuming $k_{\rm stim}^{(2)} >> k_{\rm stim}^{(3)}$ the formation of
the stimulated pump form during the first phase of stimulation depends
on $k_{\rm stim}^{(3)}$ only.  This approximation is supported by the
fact that $k_{\rm stim}^{(3)}$ converges at high calcium
concentrations in the same manner as $k^{\rm exp}_{\rm stim}$ does
(see figure \ref{fig:k_stim}). In contrast, $k_{\rm stim}^{(2)}$
diverges for high calcium concentrations (see Appendix).
\marginpar{figure \ref{fig:k_stim}} $k_{\rm stim}^{(3)}$ can be
interpreted as corresponding to the single exponential fit constant
$k^{\rm exp}_{\rm stim}$. This enables us to determine the stimulation
parameters (shown in figure \ref{fig:k_stim}).

In an analogous fashion to the stimulation case we deduce an
analytical expression like equation
\eqref{eqn:stim_stimform_estimat_1} for the relaxation scenario. The
exponent of this single exponential expression can be related to the
experimentally measured relaxation constant (see figure
\ref{fig:k_unstim}, calculation not shown). \marginpar{figure
  \ref{fig:k_unstim}}

A fit of the derived expressions for the introduced stimulation
parameters to the experimental data restricts their values.  As a
first approach we use Hill equations (see Appendix for the stimulation
case). The maximum stimulation and relaxation fit constant can be
identified with $k_3$ and $k_4$, respectively, which limits the
stimulation and relaxation rate. In contrast, the half maximum
concentration of the Hill fit is not sufficient to determine the
remaining free parameters.  Even with the knowledge of the
dissociation constant $K_2$, $K_2 = k_{-2}/k_2 = 0.5$ nM for the h2b
isoform and $K_2 = 5$ nM for the h4b isoform
\label{references}
\cite{Guerini1998,Caride2001b,Meyer1992,Elwess1997,Enyedi1991,Ba-Thein2001,Penheiter2002},
the system remains under-determined because $K$ is not precisely
known. A lower boundary for $K$ comes from the requirement of positive
reaction rates $k_2$ and $k_{-2}$ and from the value of $K_2$.  An
upper boundary for $K$ is imposed by the choice of the value for $K_4
= k_4/k_{-4}$. The parameters resulting from the Hill equation fits
are reported in table \ref{tab:para_all}. \marginpar{table
  \ref{tab:para_all}}

Attention should be drawn to the fact that Ca$^{2+}$ is bound by polar
amino acids of calmodulin and therefore the affinity represented by
$K$ strongly depends on the ionic strength of the experimental
solution \cite{Linse1991}. We assume the same value of $K$ for all
experiments, i.e. the same conditions in all experiments. Since the
value of $K$ is not exactly known for the experiments under
consideration, we assume three different values and quote the
respective parameter sets in order to show how other parameters depend
on $K$. The values used for $K$ are 0.1, 1 and 10 $\mu$M$^{-4}$ which,
according to Linse {\it et al.}, correspond to an experimental
solution with a KCl concentration of 18, 20 and 26 mM.

Using \eqref{eqn:sol_k_stim} and the parameter values determined by
the Hill equation fit we are able to calculate the calmodulin
dependence of $k_{\rm stim}$ without further assumptions.  The result
in figure \ref{fig:k_stim} (dotted line) is in qualitative agreement
with the experimental data of figure 3B in \cite{Penheiter2003}
($k_{\rm act}$).  The measurements have been carried out at a calcium
concentration of 10 $\mu$M with 0.005-0.065 $\mu$M calmodulin.  The
experimental data could be fitted by a linear equation.  The
saturation suggested in figure \ref{fig:k_stim} occurs at higher
calmodulin concentrations only. A linear fit is also suggested in
figure 4A in \cite{Caride2001a} for a range of 0.005-1.25 $\mu$M
calmodulin at a constant calcium concentration of 1 $\mu$M. However,
the linearity of the fit relies on a single data point and we consider
this result to be in agreement with the saturation predicted by our
model. We quote $k_{\rm stim}$ at a calcium concentration of 0.8
$\mu$M for the h2b isoform and at 1 $\mu$M for h4b. At both cases the
stimulation constant saturates already at low calcium concentrations,
{\it i.e.} at $\approx 0.2$ $\mu$M for h2b and at $\approx 0.4$ $\mu$M
for h4b.  The results from the Hill equation fit also allow us to
predict the calmodulin dependence of the relaxation constant $k_{\rm
  relax}$, which has not been measured so far (see figure
\ref{fig:k_unstim}).

We have seen that it is not possible to reliably determine all
stimulation parameters by a fit on the basis of Hill equations. We
therefore use a Metropolis algorithm, which is based on a
least-squares fit routine, to incorporate three sets of available data
for the stimulation constant, the relaxation constant and, in
addition, the steady-state pumping activity, which will be discussed
subsequently.  The medians and the confidence intervals in table
\ref{tab:k_fix} and \ref{tab:pump_par} are calculated from a Bootstrap
method, {\it i.e.} random data sets of the same size are drawn from
the original data set from which, using the Metropolis algorithm, the
fit parameters are calculated. Repeating this procedure 10.000 times
enables us to calculate medians and confidence intervals for the fit
parameters.  \marginpar{table \ref{tab:k_fix}} The medians of the
parameters in table \ref{tab:K_different} are determined similarly but
for different values of $K$. \marginpar{table \ref{tab:K_different}}

Note that we have derived equation \eqref{eqn:sol_k_stim} only by
considering the beginning of the stimulation. An alternative approach
to determine the parameters would be the fit of the exact model
equations to the time course of inorganic phosphate Ph$_i$ release
(whereas sub-i refers to inorganic).  Ph$_i$ emerges by hydrolyzing
ATP during pumping (see section ``Results'' and
\cite{Caride2001b,Caride1999}). However, this would not improve the
accuracy of the parameter determination since these data contain
information not only about the stimulation but also about the pumping
of the PMCA, which cannot be disentangled. Only a measurement of the
exact stimulation and relaxation behavior would provide relevant
additional information in order to get a more reliable choice in table
\ref{tab:k_fix}.



\paragraph{Stimulation dynamics} 
Based on the stimulation model reactions \eqref{buff}-\eqref{unstim}
we can simulate the dynamics of the unstimulated and stimulated PMCA
states. With the obtained isoform specific parameters differences
between the isoforms can be discussed.  The fact that the value of
$k_3$ for the h2b isoform is more than twice that of the h4b isoform
reflects faster stimulation, whereas relaxation is faster for the h4b
isoform since $k_4^{h4b} > k_4^{h2b}$.  The stimulation and relaxation
dynamics of both isoforms are shown in figure \ref{fig:dynamics}.
\marginpar{figure \ref{fig:dynamics}} In graph \ref{fig:dynamics} (a) we
start with no stimulated pump (i.e. $f_{\rm stim}(t_0 = 0) = 0$)
and expose the pump to a high calcium concentration. In the
relaxation graph \ref{fig:dynamics} (b) all pumps are stimulated
(i.e. $f_{\rm stim}(t_0=0) = 1$) and the calcium concentration
is decreased to 0.1 $\mu$M.


\subsection*{Pumping model} 

Up to this point we have described the temporal transition between
unstimulated and stimulated pump forms. In this section we will
mathematically characterize how the PMCA transports calcium ions
across the membrane depending on the intracellular calcium
concentration.  We will consider the pump cycle to be in
quasi-steady-state, {\it i.e.} the pump activity reacts immediately to
the intracellular calcium concentration. This simplification is
justified since the pump activity adapts to different calcium
concentrations within milliseconds while the stimulation and
relaxation happens on a time scale of minutes.

\paragraph{Pumping activity} 
In contrast to the Ca$^{2+}$ pump of the sarcoplasmic reticulum
(SERCA) the stoichiometry between transported Ca$^{2+}$ and hydrolyzed
ATP of the PMCA is most likely 1:1 \cite{Zylinska2000, Carafoli1991,
  Carafoli1992,Lytton1992}. The SERCA pump works with a calcium:ATP
ratio of 2:1 \cite{Carafoli1992,Lytton1992}.  Hence, the transport step of
one calcium ion through the plasma membrane becomes
\begin{equation} 
\rm P_{\rm in} \overset{r_1}{\rightharpoondown} \rm P_{\rm out} +  {Ca^{2+}}_{\rm out}. \label{pump} \\%
\end{equation}
P$_{\rm in}$ stands for a state of the pump in which one or two
calcium ions are bound and which can perform the calcium transport
process by a transformational change. P$_{\rm out}$ refers to the pump
state after the translocation step.  We do not consider the
transformational change from P$_{\rm out}$ to P$_{\rm in}$, which
closes the pump cycle.  Since ATP is assumed to be sufficiently
available in cells we treat the ATP concentration as thermodynamical
bath. The calcium transport rate $J$ in the outward direction is given
by
\begin{equation}
J = r_1 \cdot  P_{\rm in}. \label{eqn:rate}
\end{equation}

Carafoli pointed out that the activation of the ATPase by calcium and
the saturation can best be described by a Hill equation with a Hill
coefficient of two \cite{Carafoli1991}. This may be related to the
binding of 2 calcium ions at the intracellular binding site of the
pump even if only one calcium ion is translocated.  In the following
this is incorporated into the model:
\begin{eqnarray}
\rm P + Ca^{2+}  \overset{r_2}{\underset{r_{-2}}{\rightleftharpoons}} \rm  P \negmedspace \cdot  \negmedspace Ca, \label{rea:ca_bind_1} \\
\rm P \negmedspace \cdot  \negmedspace Ca + Ca^{2+}    \overset{r_3}{\underset{r_{-3}}{\rightleftharpoons}} \rm P \negmedspace \cdot  \negmedspace Ca_2. \label{rea:ca_bind_2}
\end{eqnarray}
The pump state P$_{\rm in}$ is assumed to comprise
the pumps with one and two calcium ions, {\it i.e.}  P$_{\rm in} =
\rm P \negmedspace \cdot \negmedspace Ca+ P \negmedspace \cdot
\negmedspace Ca_2$. This implies that the pump with one
calcium ion, P$\cdot$Ca, is able to perform the transport process
which is not necessarily the case but the most
general assumption (see also comments after equation
\eqref{eqn:pumping_coop}).

In the quasi-steady-state approximation of equations
\eqref{rea:ca_bind_1} and \eqref{rea:ca_bind_2} we find an expression
for $P_{\rm in}$. Using this expression the calcium transport rate
(equation \eqref{eqn:rate}) becomes
\begin{equation}
J = r_1 \frac{R_3 Ca + {Ca}^2}{R_2 R_3 + R_3 Ca + {Ca}^2}, \label{eqn:pumping_exact}
\end{equation}
with $R_2 = r_{-2}/r_2$ and $R_3 = r_{-3}/r_3$ and the pump
mass conservation $ 1 = P + P \negmedspace \cdot \negmedspace Ca + P
\negmedspace \cdot \negmedspace Ca_2$. Note that we have divided the
pump mass conservation by P$_0$ so that the different pump states
become fractions instead of concentrations. 
The sum of the fractions
of all the three pump states is of course $1$.

Fitting equation \eqref{eqn:pumping_exact} to the experimentally
measured transport rate in the absence of calmodulin yields $R_2 = 38
\,\, \mu$M and $R_3 = 0.011 \,\, \mu$M for the h2b isoform (see figure
\ref{fig:steady_state_Ca}). These values suggest a cooperative binding
of both calcium ions, {\it i.e.} the binding of the first ion is slow
but of the second fast. In the model this means $r_2 \rightarrow 0$
and $r_3 \rightarrow \infty$ while keeping $r_2 \cdot r_3$ constant,
in which case $R_2 \rightarrow \infty$ and $R_3 \rightarrow 0$ while
$R_2 R_3$ is constant.  Experiments performed by Elwess {\it et al.}
\cite{Elwess1997} or by Verma {\it et al.}  \cite{Verma1996} yield the
same sigmoidal behavior of the transport rate as found by equation
\eqref{eqn:pumping_exact} in figure \ref{fig:steady_state_Ca},
supporting the assumption of cooperative binding.
Thus \eqref{eqn:pumping_exact} is simplified to
\begin{equation}
J = r_1 \frac{{Ca}^2}{R_2 R_3 + {Ca}^2} = J_{\rm max} \frac{{Ca}^2}{{H_{1/2}}^2 + {Ca}^2} . \label{eqn:pumping_coop}
\end{equation}
In this expression the pumping activity
of the P$\cdot$Ca state which has only one calcium ion bound is neglected. 
The high
cooperativity prevents us from deciding whether this state is
contributing to the total pump activity or not.

Equation \eqref{eqn:pumping_coop} has the form of a Hill
equation with Hill coefficient $n=2$.  $r_1$ can be identified with
the maximal pump rate $J_{\rm max}$.  Since the maximum rate of the
calcium pumping is limited by $r_1$ in equation \eqref{eqn:pumping_coop},
$r_1 = J_{\rm max}$ stands for the saturating
properties of the pumping process. $H_{1/2} = \sqrt{R_2 R_3}$ is
the half activation concentration and a measure for
the affinity of the pump to calcium. We will use equation
\eqref{eqn:pumping_coop} for all further investigations in this paper.

So far we have only looked at the calcium-dependent pump activity in
the absence of calmodulin without considering pump stimulation.  The
activity in the unstimulated state corresponds to the base activity in
the absence of calmodulin. It is very likely that the interaction with
calmodulin causes the increase in pumping activity since it is
well-known that calmodulin increases both the affinity for calcium as
well as the maximal activity and both effects occur during stimulation
\cite{Penheiter2003}. The transition dynamics is given by the
stimulation model in equations \eqref{buff}-\eqref{unstim}. Hence, the
total pumping rate of a system becomes
\begin{equation}
J = f_{\rm unstim}\cdot \frac{{J_{\rm max}} \cdot Ca^2}{ {H_{1/2}}^2 + Ca^2} + f_{\rm stim} \cdot \frac{{J_{\rm max}^*} \cdot Ca^2}{ {H_{1/2}^*}^2 + Ca^2},
\label{eqn:J_total}
\end{equation}
where $f_{\rm unstim}$ and $f_{\rm stim}$ are the fractions of
unstimulated and stimulated pump with $f_{\rm unstim} = \frac{P +
  P_X}{P_0}$ and $f_{\rm stim} = \frac{P^* + P_X^*}{P_0}$.

One input parameter of the stimulation model is the total pump
concentration $P_0$. This concentration $P_0$ is taken from Caride
{\it et al.} \cite{Caride2001b} to be $0.005 \, \mu \rm M$.  Assuming
a sufficiently large amount of pumps, the total pump concentration is
irrelevant for the pumping model since we consider fractions of pumps
in different states only. In contrast in the stimulation model the
proportion between the total calmodulin concentration, $M_0$, and the
total pump concentration, $P_0$, is significant. If for example the
amount of PMCA's exceeds the number of available calmodulin proteins
not all pumps could be stimulated.

For simulations the surface density of PMCA's has to be transfered
into the concentration $P_0$. This implies the neglection of diffusion
of the calcium-calmodulin complex to and from the pump, which is
justified for rather small compartments only. For larger
compartments a space resolving model with a local pump concentration
has to be used.


\paragraph{Pumping parameter determination with steady-state pump activity} 

By setting all differential equations (equations \eqref{eqn:diff_p},
\eqref{eqn:diff_pc*} and \eqref{eqn:diff_p*}, see Appendix) of the
stimulation model (reactions \eqref{buff}-\eqref{unstim}) to zero the
steady-state distribution of P, P$_{\rm X}$, P$_{\rm X}^*$ and P$^*$
is found. This determines the asymptotic steady-state fractions
$f_{\rm unstim}$ and $f_{\rm stim}$ in equation \eqref{eqn:J_total}.
Using this distribution we calculate the asymptotic pumping activity
in dependence of the calcium and calmodulin concentration.
Many measurements of the steady-state pump activity in the presence
and the absence of calmodulin have been performed \cite{Caride2001b,
  Ba-Thein2001,Verma1996, Hilfiker1994, Adamo1998}.  Fitting equation
\eqref{eqn:J_total} to one of these experiments yields $J_{\rm max}$,
$J_{\rm max}^*$, $H_{1/2}$ and $H_{1/2}^*$. In figure
\ref{fig:steady_state_Ca} we use the steady-state activity
measurements done by Caride {\it et al.}  \cite{Caride2001b}.
\marginpar{figure \ref{fig:steady_state_Ca}}

With $f_{\rm unstim} = 1$ and $f_{\rm stim} = 0$ in equation
\eqref{eqn:J_total} we fit the steady-state activity curve in the
absence of calmodulin which supplies $J_{\rm max}$ and $H_{1/2}$ in
equation \eqref{eqn:pumping_coop}.  Equation \eqref{eqn:J_total} is
used to fit the data in the presence of calmodulin (see dashed line in
figure \ref{fig:steady_state_Ca}).  We determine $J_{\rm max}^*$ and
$H_{1/2}^*$ with that fit, whereby the steady-state fractions $f_{\rm
  unstim}$ and $f_{\rm stim}$ are determined by the stimulation model.
These fractions respect the fact that at lower calcium concentrations
in the presence of calmodulin no stimulated pump form is available
($f_{\rm stim} \simeq 0$). This can be seen by inspecting the
dashed-dotted line in figure \ref{fig:steady_state_Ca}, which is
plotted with the artificial assumption that only stimulated pump form
($f_{\rm stim} = 1$) is present for all calcium concentrations.  The
gap between this virtual line (dashed-dotted) and the fit of equation
\eqref{eqn:J_total} (dashed line) at low calcium concentrations is
determined by the choice of the stimulation parameter set of table
\ref{tab:k_fix}.  Although we determine $J_{\rm max}^*$ and
$H_{1/2}^*$ with that fit, changes in $J_{\rm max}^*$ and $H_{1/2}^*$
only alter the slope and the saturation value whereas the gap at low
calcium concentrations is not influenced by these changes: $J_{\rm
  max}^*$ and $H_{1/2}^*$ do not alter the low calcium concentration
limit.  Note that the fit of the case with calmodulin being present,
and therefore the determination of $J_{\rm max}^*$ and $H_{1/2}^*$, is
accomplished with the Bootstrap method using the Metropolis fit
algorithm that also takes into account the stimulation and relaxation
constant data sets.  The pumping parameters obtained from this routine
are summarized in table \ref{tab:pump_par}. The impact of stimulation
by calmodulin on PMCA function, namely to increase both the affinity
to calcium $H_{1/2}$ and the maximum pump rate $J_{\rm max}$, can be
affirmed.  \marginpar{table \ref{tab:pump_par}}


The obtained parameters $J_{\rm max}$ and $J_{\rm max}^*$ refer to the
investigated system, {\it i.e.} they are determined by the ensemble of
investigated pumps. In contrast to the pump rate, the affinity to
calcium expressed by $H_{1/2}$ or $H_{1/2}^*$ is a universal parameter
since it does not depend on the surface density.  Knowing the
expression level of PMCA's within the system and the average protein
mass one can use $J_{\rm max}$ and $J_{\rm max}^*$ to calculate the
maximum unstimulated $J_{\rm single}$ and stimulated pumping rate
$J^*_{\rm single}$ of a single pump respectively, which is then a
system-independent parameter characteristic for the isoform studied.
Caride {\it et al.} achieved with the baculovirus expression system an
amount of 5\% PMCA's of total membrane protein (private
communication). With this data and with an overall average protein
mass of 130 kDa \cite{Carafoli1992} we deduced a turnover rate of 10.4
Hz out of $J_{\rm max}^* = 0.24\,\mu \rm mol/(mg \cdot min)$ for the
h2b isoform and 31.7 Hz out of $J_{\rm max}^* = 0.73\,\mu \rm mol/(mg
\cdot min)$ for the stimulated h4b rate.


With the use of the pumping parameters and the stimulation rate
constants (table \ref{tab:k_fix}) we can calculate the calmodulin
dependence of the steady-state activity (see figure
\ref{fig:steady_state_CaM}).  \marginpar{figure
  \ref{fig:steady_state_CaM}} The displayed fractional steady-state
activity $f$ is a universal description, {\it i.e.} it is not
dependent on the surface density.
%
%
The quantitative behavior of our calculation is confirmed by
experimental data of Penheiter {\it et al.} \cite{Penheiter2002}
(compare our figure \ref{fig:steady_state_CaM} with figure 3 on page
17730 in this publication).  Note that we calculated this steady-state
activity out of the available data without further fitting routines.



\section*{Results}

\paragraph{Comprehensive PMCA dynamics}

With the knowledge of the stimulation parameters from table
\ref{tab:k_fix} and the pumping properties in table \ref{tab:pump_par}
we are now able to simulate the time dependent behavior of the calcium
pumping rate of the PMCA including the stimulation.

In a corresponding experiment \cite{Caride2001b} tissue with PMCA
pumps of isoform h2b was exposed consecutively to different calcium
concentrations. The time course of Ph$_i$ produced is shown in figure
\ref{fig:dyn_activity} (a) (crosses).  It is rather likely that during
a single turnover of the PMCA one ATP is hydrolyzed as well as one
calcium ion transported \cite{Zylinska2000, Penniston1998,
  Carafoli1991, Carafoli1992}. Therefore the rate of Ph$_i$ production
directly corresponds to the calcium pumping rate.
\marginpar{figure \ref{fig:dyn_activity}} Utilizing equation
\eqref{eqn:J_total}, we can calculate and sum up the time course of
Ph$_i$ production. With the use of the universal parameter sets we
adapt our simulation (full line in figure \ref{fig:dyn_activity}) to
this experiment by adjusting the unknown surface density of PMCA's.
This yields an expression level of 9.8 \% PMCA's of total membrane
protein, which is comparable to the quoted 5 \% (see before).  Note
that we have added a constant base level rate with respect to the
non-zero Ph$_i$ production rate in the absence of calcium ($Ca=0 \,
\mu$M) in figure \ref{fig:dyn_activity} (a), which has to be related
to a different source within the investigated tissue.  In fact, in the
absence of calcium no ions can be transported, {\it i.e.}  no ATP can
be hydrolyzed, no Ph$_i$ produced.
Note that during the second exposure to 0.5 $\mu$M calcium the
measured activity significantly decreases compared to the first time
at high calcium concentration, {\it i.e.} even under the same
experimental conditions the Ph$_i$ production rate changes during the
experiment. This might be related to dwindling resources such as ATP
or the increasing significance of ATP and calcium diffusion. Such
effects naturally cannot be reproduced by the present model.

After 300 seconds at 0.5 $\mu$M calcium, during which a PMCA fraction
is stimulated, the concentration was decreased to 0.05 $\mu$M calcium
for different durations and raised again to high calcium
concentration. The turnover rate at the end of the second low calcium
exposure $J^{(\rm second)}_{0.05}$ is, due to the gradual decay of
stimulated pump form, dependent on the duration of the low calcium phase.
The fraction of change in $J_{0.05}$, $\frac{J^{(\rm second)}_{0.05} -
  J^{(\rm first)}_{0.05}}{J^{(\rm first)}_{0.5}-J^{(\rm
    first)}_{0.05}}$ is shown in figure \ref{fig:dyn_activity} (b).
The temporal behavior
of pump function is quantitatively reproduced.


\section*{Discussion}

Our combined model of stimulation and pumping is able to reproduce the
PMCA behavior. Therefore the assumptions of the stimulation model
could serve as a possible explanation for the underlying biological
steps. An essential element of the model is the separation of the
attachment of the calmodulin-calcium complex to the pump (reaction
\eqref{bind}) and the stimulation, which occurs in an additional step
(reaction \eqref{stim}). The convergence of the stimulation constant
$k^{\rm exp}_{\rm stim}$ at high calcium concentrations in figure
\ref{fig:k_stim} provides strong evidence for that assumption.  The
translocation step in reaction \eqref{pump} has the same
characteristics.  These steps can be seen as internal conformational
changes where their velocity cannot be further increased by higher
concentrations.  In contrast to reaction \eqref{stim} which is
confined by the total available pump concentration $P_0$ and $k_3$,
the velocity of reaction \eqref{bind} can be arbitrarily increased
with higher calcium or calmodulin concentrations.  Similarly the
saturation of the relaxation constant $k^{\rm exp}_{\rm unstim}$ at
low calcium concentrations calls for a rate limiting reaction during
relaxation.  Regarding the relaxation pathway in this context yields
two steps which could meet this property, {\it i.e.}  forward reaction
\eqref{detach} and reaction \eqref{unstim}.  The Metropolis fit
algorithm provides similar outcomes for both possibilities. In general
the routine delivers $k_4 < k_5$, {\it i.e.}  reaction \eqref{detach}
limits the relaxation rate. The experimental data could also be
reproduced assuming $k_4 > k_5$, which corresponds to the irreversible
relaxation being the limiting step. However, figure 4b in
\cite{Penheiter2003} provides clear evidence that the limiting step is
related to the dissociation of calmodulin, hence step \eqref{detach}
in our model.

The separation between stimulation (reactions \eqref{bind} and
\eqref{stim}) and relaxation (reactions \eqref{detach} and
\eqref{unstim}) enables us to describe the different temporal
behaviors independently. This is required since for example the
isoforms h2b and h4b display opposite stimulation and relaxation
behavior: The h2b isoform is stimulated more quickly than h4b but
relaxes more slowly under the same conditions.  Assuming the
stimulation reaction \eqref{stim} would become reversible, the
formation and degradation of stimulated pump form $\rm P_{\rm X}^*(t)$
could be described together by the first three reactions \eqref{buff},
\eqref{bind} and \eqref{stim} without the use of reactions
\eqref{detach} and \eqref{unstim}. In such a case the calcium and
calmodulin dependent stimulation and relaxation velocity could not be
adjusted autonomously which is required for the reproduction of the
experimental data. The formal introduction of two different pathways
may be interpreted as different underlying transformational changes.

Comparing our parameters with those of the model proposed by Penheiter
{\it et al.}  in \cite{Penheiter2003} reveals strong agreement. These
authors argue that the preferred stimulation route in their branched
model is the binding of the X$_4$ complex to the unstimulated (closed,
in their terminology) conformation and the subsequent stimulation
(opening) of the pump.  This corresponds to our single stimulation
step.  The second route assumes a stimulated (open) state of the pump
in the absence of the X$_4$ complex which would be stabilized by the
binding of the calmodulin-calcium complex.  This branch had to be
introduced in \cite{Penheiter2003} in order to describe the initial
increase in fluorescence.  As in our model, the conformational change
limits the stimulation. The limiting steps of the Penheiter {\it et
  al.}  model are determined by $k_1^{\rm p}$ and $k_4^{\rm p}$, where
$k_4^{\rm p}$ describes the preferred stimulation route (here p refers
to reaction rates in the Penheiter model).  The value for $k_4^{\rm p}
= 0.034$ 1/s is similar to our $k_3$ (see h4b isoform in table
\ref{tab:para_all} and \ref{tab:k_fix}). Their and
our model confirm the known affinity of the h4b isoform for the
calmodulin-calcium complex
\cite{Meyer1992,Caride1999,Enyedi1991,Penheiter2002}.

Since we have based the parameter deduction on inorganic phosphate,
Ph$_i$, release measurements and the Penheiter model relies on
fluorescence measurements based on the binding of calmodulin to the
pump, the agreement of the parameters justifies both approaches.
However, the reduction of the Penheiter model to the stimulation
pathway ending in a stimulated pump with isomerized CaM protein
(TA-CaM-PMCA$_0^*$) might not account for the competing stimulation
and relaxation processes of the pump. This fact together with the
branched model ansatz could be the reason for slightly different
reaction rates.

As already mentioned, the choice of $k_i$ (i$=\pm 2,3,\pm 4,5$) and
therefore the choice of $K$ has a sensitive effect on the stimulation
and relaxation graphs in figure \ref{fig:k_stim} and
\ref{fig:k_unstim}. Furthermore the steady-state graphs in figure
\ref{fig:steady_state_Ca} are strongly influenced by these parameters.
The attempt to reproduce all experimental data for both isoforms on
the basis of our model by using a single value of $K$ leads us to the
following conclusion: The affinity of the PMCA for the X$_4$ complex
is altered during the stimulation process.  The unstimulated pump P
exhibits a lower affinity than the stimulated pump P$^*$ (compare $K_2
= k_{-2}/k_2$ to $K_4 = k_4/k_{-4}$).  For both isoforms the value of
$K_4$ is thousandfold lower compared to $K_2$ ($K_4 = 0.0008$ nM for
h2b, $K_4 = 0.002$ nM for h4b). Within the framework of our model the
assumption $K_2 = K_4$ fails to adequately reproduce all experimental
data. In the case of the isoform h4b this assumption does not allow a
sufficient amount of PMCA to become stimulated in the steady-state
calculation.  In principle the experimental data in figure
\ref{fig:steady_state_Ca} could still be fitted by adjusting
$H_{1/2}^*$ and $J_{\rm max}^*$, but this would contradict the
statement that $J_{\rm max}$ is about 20\% of $J_{\rm max}^*$ for the
h4b isoform \cite{Caride2001b,Elwess1997,Ba-Thein2001,Penheiter2002}.
Note that this result is indirectly supported by the observation that
the Ca$^{2+}$/calmodulin-dependent protein kinase II (CaMKII) target
protein has a thousandfold higher affinity to the X$_4$ complex in the
stimulated state, an effect denoted as calmodulin trapping
\cite{Hudmon2002}.
  
The fit results in table \ref{tab:K_different} show that apart from
$k_2$ and $k_5$ the parameters remain relatively constant for
different values of $K$. Hence, an exact determination of $K$ under
the specific experimental conditions could improve the determination
of $k_2$ and $k_5$ but would not change the remaining values of the
stimulation, the pumping parameters or the main conclusions. Note,
that for all of the three assumed values, {\it i.e.}  $K=0.1$ , 1 and
10 $1/ \mu$M$^4$, the experimental data of both isoforms could be
reproduced with equal accuracy.


We have also investigated a different approach for PMCA relaxation.
In this scenario only calcium dissociates from the complex P$^*_{\rm
  X}$ in step \eqref{detach}'. The following reaction \eqref{unstim}'
comprises the relaxation and the detachment of calmodulin. Though the
reproduction of the available experimental data is ensured with this
approach, there are two aspects which distinguish it from the present
model.  First, the analytical expression for the relaxation constant
becomes independent of calmodulin due to the irreversibility of step
\eqref{unstim}'.  It is, therefore, impossible to predict the
calmodulin dependence of the relaxation rate as done in figure
\ref{fig:k_unstim}. Second, similar to the change of the pump affinity
for the X$_4$ complex in the present model, the affinity of calmodulin
for calcium would be different depending on whether calmodulin is free
(reaction \eqref{buff}) or bound to the PMCA pump (reaction
\eqref{detach}'). We can find experimental support for both scenarios.
As previously quoted, CaMKII is known to change its affinity for the
calmodulin-calcium complex \cite{Hudmon2002}, whereas Olwin {\it et
  al.} report a change of the dissociation constant between CaM and
calcium over two orders of magnitude in the presence and in the
absence of a target protein, {\it i.e.} in their case the rabbit
skeletal muscle myosin light chain kinase \cite{Olwin1984}. No
experimental evidence is known to us whether calcium or the whole
X$_4$ complex detaches first from the pump. However, since we treat
the balance between free X$_4$, calmodulin and calcium in
quasi-steady-state our approach presented here is the more general one
since reaction \eqref{detach} makes only the statement that the
detachment of calmodulin limits the relaxation rate but does not
restrict in which order the components dissociate.

In the description of the pumping behavior we have incorporated the
experimental observations that although the translocation step
involves only one calcium ion (indicated by the 1:1 stoichiometry of
ATP to transported calcium) the pump binds two Ca$^{2+}$ ions with
high affinity
\cite{Zylinska2000,Penniston1998,Carafoli1991,Carafoli1992}.  The
experimental data provide strong evidence for the assumption that the
binding of the two calcium ions to the intracellular binding site of
the pump is highly cooperative.  The deduced pumping rate expression
is in accordance with these observations.  The high cooperativity
prevents us from determining whether only P$\cdot$Ca$_2$ or also
P$\cdot$Ca can perform the transport process of one calcium ion.  Our
general ansatz could in principle account for recent findings of
Guerini {\it et al.}  in which the Ca$^{2+}$-dependent ATPase activity
is fitted by a Hill function with Hill coefficient 1
\cite{Guerini2000}. However, the fit to the available data (see figure
\ref{fig:steady_state_Ca}) suggests a Hill coefficient 2.

Our aim to characterize pump isoforms by different universal parameter
sets is achieved for the h2b and h4b isoforms. We assume the same
functionality for all isoforms, hence these parameters express
relevant differences in the stimulation and pumping behavior. With the
concept of providing universal parameter sets we claim that the sets
are applicable to other simulations including hPMCA2b or hPMCA4b
isoforms.

Out of the universal parameters of the h4b isoform we can
calculate an accurate prediction of the calmodulin dependence of the
steady-state activity. 
Using the expression level
and the overall average protein mass we calculated single pump
turnover rates.  Based on a measurement of Elwess {\it et al.} in 1997
\cite{Elwess1997} and using an expression level of 0.2 \%
we computed a turnover rate of 8.5 Hz for the rPMCA2b
isoform. Our calculated single turnover rates (table \ref{tab:pump_par})
are in the same range as those of Blaustein who indicated a turnover
of $\approx$ 30 Hz without specifying the isoform
(\cite{Juhaszova2000} and private communication). Adamo {\it et al.}
made direct measurements of the turnover rate and determined it to be
33 Hz for the erythrocyte PMCA \cite{Adamo1988}.  However, the
calculated turnover rates, even being in the appropriate range, may 
differ from the presented values
since the expression levels of the
pump can not be determined with much accuracy, and factors such as
other proteins and lipids interacting with the pump in the biological
membrane environment can have substantial effects.

For comparison, the turnover rate of the Sarcoplasmic Reticulum
Ca$^{2+}$-ATPase (SERCA) lies within the same range. The SERCA1,
SERCA2a and SERCA3 isoforms transport calcium with a turnover rate of
$\approx 10$ Hz, whereas the the SERCA2b isoform exhibits a rate of
$\approx 5$ Hz \cite{Lytton1992}. Note that in contrast to the PMCA
the SERCA carries two calcium ions per pump cycle reflected
in the Ca$^{2+}$:ATP stoichiometry of 2:1
\cite{Carafoli1992,Lytton1992}.

The isoform specific universal set of parameters may be applied to
different scenarios by incorporating the system specific expression
levels of PMCA. However, alterations of the stimulation rate constants
may occur under modified experimental conditions. Since we consider
proteins with polar binding sites we cannot exclude a change of
dynamical properties at different ionic strengths of the experimental
solutions. Corresponding experiments could elucidate the importance of
such effects. Also our theoretical proposition of the calmodulin
dependence of the stimulation and the relaxation constant which
differs from the linear behavior found by Caride {\it et al.}  calls
for further experimental investigations.


\section*{Acknowledgment}

This work has been completed during support of Michael Graupner by a
French Government scholarship in conjunction with a DAAD supplement
and during support of M. Meyer-Hermann by a Marie Curie Intra-European
Fellowship within the Sixth EU Framework Programme.  We are indebted
to Ariel Caride for kindly providing us the data and fruitful
communications. We also would like to thank Jakob Schweizer and
Nicolas Brunel for many discussions.  Furthermore we want to thank
Alexander Roxin for revising the manuscript.


\section*{Appendix}

\paragraph{Stimulation model}
By using the Law of Mass Action the stimulation model
\eqref{buff}-\eqref{unstim} is rewritten as a set of differential equations for
concentrations
\begin{eqnarray}
\frac{dP}{dt} & = & - \frac{k_2 P (M_0 - P_X - P_X^*)}{1+1/(K Ca^4)} + k_{-2} P_X + k_5 ( P_0 - P - P_X - P_X^*) \label{eqn:diff_p}, \\
\frac{dP_X}{dt} & = & \frac{k_2 P (M_0 - P_X - P_X^*)}{1+1/(K Ca^4)} - (k_{-2} + k_3) P_X \label{eqn:diff_pc*}, \\
\frac{dP_X^*}{dt} & = & k_3 P_X - k_4 P_X^* + k_{-4} (P_0 - P - P_X - P_X^*)\frac{M_0 - P_X - P_X^*}{1+1/(K Ca^4)} \label{eqn:diff_p*}.
\end{eqnarray} 
Pump conservation $P_0 = P + P_X + P_X^* + P^*$ where $P_0$ is the
total pump concentration and calmodulin conservation $M_0 = M + X_4 +
P_X + P_X^*$ with $M_0$ as total calmodulin concentration is
respected.  We solve this system of coupled ordinary differential
equations of first order with a fourth-order Runge-Kutta method. This
has been implemented in a C++ program.

\paragraph{Hill fit of the stimulation constant}

The data from \cite{Caride2001b} can be fitted by the Hill equation
\begin{equation}
  k_{\rm stim}^{\rm exp}(Ca) = \frac{k_{\rm stim}^{\rm max}\cdot Ca^4}{{Ca_{\rm stim}^{(1/2)}}^4 + Ca^4}, \label{eqn:k_stim_fit_Ca}
\end{equation}
with the maximal stimulation constant $k_{\rm stim}^{\rm max}$ and the
half maximum concentration $Ca_{\rm stim}^{(1/2)}$.  According to
\eqref{eqn:sol_k_stim}, $k_{\rm stim}^{(3)}$ converges to $k_3$ for
large calcium or calmodulin concentrations (see below), hence $k_3$
confines the maximum stimulation rate and can be identified with
$k_{\rm stim}^{\rm max}$. This reflects the fact that equation
\eqref{stim} limits the stimulation rate.  By setting the right hand
side of equation \eqref{eqn:sol_k_stim} equal to $k_3/2$, an
expression for the Ca$^{2+}$ concentration at half maximal stimulation
is derived
\begin{equation}
Ca_{\rm stim}^{(1/2)}  =  \sqrt[4]{\frac{k_{-2} + k_3/2}{k_2 \cdot K \cdot M}}. \label{eqn:Ca_stim_1/2}
\end{equation}
Note that the calmodulin concentration $M$ is kept constant.

In a similar manner, the experimental relaxation constants can be
fitted by an inverse Hill equation.  Note that we expect the
dependence of the stimulation and relaxation constant on calmodulin to
be fitted by a Hill equation with Hill coefficient 1.


\paragraph{Limes of equation \eqref{eqn:sol_k_stim}}
We consider the limes of $k_{\rm stim}^{(2/3)}$ for $X_4 \rightarrow$
$\infty$ in the case of real and positive rate constants $k_2$,
$k_{-2}$ and $k_3$.
\begin{align}
\underset{X_4 \rightarrow \infty}{\lim} k_{\rm stim}^{(2/3)} & = \underset{X_4 \rightarrow \infty}{\lim}   \frac{k_{-2} + k_3 + k_2 X_4}{2} \pm \sqrt{\left(\frac{k_{-2} + k_3 + k_2 X_4}{2}\right)^2 - k_2 k_3 X_4}  \nonumber \\
& = \underset{X_4 \rightarrow \infty}{\lim}  \frac{k_{-2} + k_3 + k_2 X_4}{2} \left( 1 \pm \sqrt{1-\frac{4 k_2 k_3 X_4}{(k_{-2} + k_3 + k_2 X_4)^2}} \right)  
\end{align} 
For $k_{\rm stim}^{(2)}$ this expression diverges, whereas 
we find
$\underset{X_4 \rightarrow \infty}{\lim} k_{\rm stim}^{(3)} = k_3$
using the rule of de l'Hospital.

\bibliographystyle{apsrev_neu}
\bibliography{pmca_model}


\clearpage
\section*{Tables}

\begin{table}[h] 
\[
\begin{array}{|l|c|c|l|c|} \hline
{\bf parameter} & {\bf h2b} & {\bf h4b}  &  {\bf unit}  & {\bf source} \\ \hline \hline
K              & \multicolumn{2}{c|}{>0.029} & 1/\mu \rm M^4   & \mbox{see text}  \\ \hline
k_3            & 0.055     & 0.024    & 1/\rm s             & k_{\rm stim}^{\rm max}    \\ \hline
k_4            & 0.015    & 0.0347   & 1/\rm s  &  k_{\rm relax}^{\rm exp} \\ \hline
\end{array} 
\]
\caption{{\bf Stimulation parameters and ranges determined by Hill
    equation fits} \newline
Pump specific rate constants and ranges determined
   by
   fits of the
 Hill equation to the experimental data from Caride {\it et al.} (2001) \cite{Caride2001b}.}
\label{tab:para_all}
\end{table}

\begin{table}[h]
\[
\begin{array}{|l|c|c|c|c|l|} \hline
{\bf parameter} & {\bf h2b} & {\bf c.i. (68 \%)} & {\bf h4b}  & {\bf c.i. (68 \%)} & {\bf unit}  \\ \hline \hline
k_2            & 1.99  & 1.92 ... 8.91 & 0.096 & 0.088 ... 0.168 & 1/(\rm \mu M\cdot
s)   \\ \hline 
k_{3}         & 0.056  & 0.053 ... 0.058  &  0.023 & 0.021 ... 0.025  & 1/\rm s               \\ \hline 
k_4            & 0.016 & 0.015 ... 0.052 & 0.035 & 0.035 ... 0.147 & 1/\rm s
  \\ \hline 
k_{-4}         & 20001 & 19988 ... 20010 & 19999 & 19978 ... 20017 & 1/(\rm \mu M\cdot s)  \\ \hline 
k_{5}         & 0.12 & 0.02 ... 0.35 & 0.84 & 0.14 ... 0.97 & 1/\rm s   \\ \hline 
\end{array} 
\]
\caption{ {\bf Determination 
    of stimulation parameters with a Metropolis algorithm} 
\newline
The medians and the confidence
intervals (c.i.) of 68 \% are calculated with 
a Bootstrap method combined with
a Metropolis algorithm using experimental data for 
the stimulation constant, relaxation
constant and steady-state pumping activity (see
text for more details). $K$ is chosen to be 1 $\mu$M$^{-4}$. $k_{-2}$
can be calculated from $k_2$ using the known dissociation constant $K_2 = k_{-2}/k_2$
(see text).}
\label{tab:k_fix}
\end{table}
\begin{table}[h]
\[
\begin{array}{|l|c|c|c|c|c|c|c|} \hline
{\bf parameter} & \multicolumn{3}{c|}{{\bf h2b}} & \multicolumn{3}{c|}{{\bf h4b}} & {\bf unit}  \\  \hline
K             & 0.1   & 1     & 10    &  0.1  & 1 & 10 & 1/\mu \rm M^4  \\ \hline \hline
k_2           & 24.8  & 1.99  & 0.20  &  1.46 & 0.096 & 0.0093 &  1/(\rm \mu M\cdot
s)   \\ \hline 
k_{3}         & 0.058 & 0.056 & 0.056 & 0.023 & 0.023 & 0.023  & 1/\rm s               \\ \hline 
k_4           & 0.015 & 0.016 & 0.018 & 0.036 & 0.035 & 0.035  & 1/\rm s
  \\ \hline 
k_{-4}        & 19991 & 20001 & 20001 & 19993 & 19999 & 19998   & 1/(\rm \mu M\cdot s)  \\ \hline 
k_{5}         & 0.019 & 0.12  & 0.68  & 0.06  & 0.84  & 7.4     &  1/\rm  s   \\ \hline 
J_{\rm max}^* & 0.24  & 0.24  & 0.24  & 0.73  & 0.73  & 0.76    &  \frac{\mu\rm mol}{\rm mg
  \cdot min} \\ \hline
H_{1/2}^*     & 0.44  & 0.45  & 0.45  & 0.56  & 0.57  & 0.54    &  \mu \rm M \\  \hline
\end{array} 
\]
\caption{{\bf Parameters at different values of $K$} \newline
See table \ref{tab:k_fix}, \ref{tab:pump_par} and the text for
more details concerning the determination of these values.}
\label{tab:K_different}
\end{table}


\begin{table}[h]
{\small
\[
\begin{array}{|l|c|c|c|c|l|c|} \hline
{\bf parameter} & {\bf h2b} & {\bf c.i. (\% 68)}  & {\bf h4b}  & {\bf
  c.i. (\% 68)} & {\bf unit}  & {\bf characteristic} \\ \hline \hline
J_{\rm max}       & 0.116 \, ( 48 \%\, of J_{\rm max}^*) & \pm 0.008 &
0.148 \, ( 20 \%\,  of J_{\rm max}^*) & \pm 0.018 &  \frac{\mu\rm mol}{\rm mg \cdot min} & \mbox{system specific} \\ \hline
J_{\rm single}       & 5.0 & \pm 0.3 & 6.4 & \pm 0.8 & \rm Hz & \mbox{universal} \\ \hline
J_{\rm max}^*     & 0.24 & 0.23 ... 0.26 & 0.73 & 0.68 ... 0.75 &   \frac{\mu\rm mol}{\rm mg \cdot min} & \mbox{system specific} \\ \hline 
J_{\rm single}^*     & 10.4 & 10.0 ... 11.3 &  31.7 & 29.5 ... 32.5 & \rm Hz & \mbox{universal} \\ \hline 
H_{1/2}       & 0.63   & \pm 0.07  & 1.45  & \pm 0.26 &   \mu \rm M      &  \mbox{universal}  \\ \hline 
H_{1/2}^*     & 0.45  & 0.29 ... 0.53  & 0.57  & 0.46 ... 0.62 &  \mu \rm M & \mbox{universal} \\ \hline
\end{array} 
\]
}
\caption{{\bf Pumping parameters} \newline
Parameters of the steady-state pump activity determined using
experimental data of Caride {\it et al.} 2001 \cite{Caride2001b}. $K$
is chosen to be 1 $\mu$M$^{-4}$. The activity values in the absence
of calmodulin are determined by 
a gnuplot fit to the experimental
data. The medians and the  confidence
intervals (c.i.) of $J_{\rm max}^*$ and $H_{1/2}^*$ are derived 
from a
Bootstrap method combined with the Metropolis algorithm (see
text for more details).}
\label{tab:pump_par}
\end{table}
\clearpage

\section*{Figure legends}
\paragraph{Figure 1}
$k_{\rm stim}^{(3)}$ is plotted against calcium at
constant calmodulin concentration of 0.117 $\mu$M
for isoforms h2b and h4b according to equation
\eqref{eqn:sol_k_stim} (full lines).
The dotted lines
represent the $k_{\rm stim}^{(3)}$ dependence on calmodulin at fixed
calcium concentration of 0.8 $\mu$M for the h2b and 1 $\mu$M for the
h4b isoform. The stimulation rate $k_{\rm stim}^{\rm exp}$ (squares)
has been measured at various calcium concentration with constant
calmodulin concentration of 0.117 $\mu$M (with kind permission of
Caride \cite{Caride2001b}).

\paragraph{Figure 2}
The relaxation constant $k_{\rm relax}$ is plotted for
varying calcium (full lines) and calmodulin (dotted lines)
concentrations where calmodulin ($M = 0.117 \, \, \mu$M for both isoforms)
and calcium ($Ca = 0.25 \, \, \mu$M for h2b; $Ca = 0.3 \, \, \mu$M for h4b) are
kept constant, respectively. 
Corresponding measurements
at $M =0.117 \, \, \mu$M are shown (squares).
With kind permission of Caride \cite{Caride2001b}.

\paragraph{Figure 3}
The dynamics of the
fractions $f_{\rm unstim} = \frac{P + P_X}{P_0}$ (full lines for
h2b, dashed-dotted lines for h4b) and $f_{\rm stim} = \frac{P^* +
  P_X^*}{P_0}$ (dashed lines for h2b, dotted lines for h4b) are shown
during stimulation (panel (a)) and relaxation (panel (b)). The
total calmodulin concentration is 0.117
$\mu \rm M$ and $P_0 = 0.005\, \mu \rm M$. Note that
the formation and degradation of stimulated and
unstimulated pump form depend on the available calcium and
calmodulin concentration. The present calcium concentration is 0.5
$\mu \rm M$ during stimulation (a) and 0.1 $\mu \rm M$ during
relaxation (b).

\paragraph{Figure 4}
Steady-state calcium dependent pump activity of both isoforms. The
measurements in the absence (triangles) and in the presence (circles)
of $0.117 \, \mu \rm M$ calmodulin have been kindly provided by Caride
\cite{Caride2001b}. The total pump concentration $P_0$ is taken from
\cite{Caride2001b} to be $0.005 \, \mu \rm M$.  
Without calmodulin (full line) the fit of
equation \eqref{eqn:pumping_exact} to the data displays no difference
to the fit of the simplified equation \eqref{eqn:pumping_coop} which
presumes a cooperative binding.  Therefore, only the fit of equation
\eqref{eqn:pumping_coop} is shown.
The dashed line shows
the realistic case and displays  $f_{\rm stim} \simeq 0$ for low
calcium concentrations. 
The dashed-dotted line
depicts the pump activity with the assumption that only stimulated
pump form is present at all calcium concentrations, {\it i.e.} $f_{\rm
  stim} = 1$ and $f_{\rm unstim} = 0$. 

\paragraph{Figure 5}
The fractional calmodulin dependent steady-state pump activity $f = (J
- J_{\rm max})/(J^*_{\rm max} - J_{\rm max})$ of the h4b isoform at
$Ca = 0.7 \,\, \mu$M is shown. $J_{\rm max}$ is the activity in the
absence of calmodulin and $J^*_{\rm max}$ is the steady-state activity
in the presence of saturating calmodulin, both at the accordant
calcium concentration of 0.7 $\mu$M. The calculation has been done
with $P_0 = 0.005 \,\, \mu$M.

\paragraph{Figure 6}
Figure (a) shows the theoretical (full line) and the experimentally
measured (crosses) time course of Ph$_i$ production during different
calcium concentration exposures, shown on the top of the panel. The
present calmodulin concentration was 0.117 $\mu$M. Figure (b) shows
the fraction of change in $J_{0.05}$, $\left( \frac{J^{(\rm
      second)}_{0.05} - J^{(\rm first)}_{0.05}}{\Delta J} \right)$ as
a function of time at low calcium (0.05 $\mu$M), experimentally
measured (crosses) and theoretically predicted (full line). Both
experiments were performed with the h2b isoform (with kind permission
of Caride \cite{Caride2001b}).

\clearpage
\section*{Figures}
\begin{figure}[h]
\centering
\subfigure[h2b isoform]{\epsfig{figure=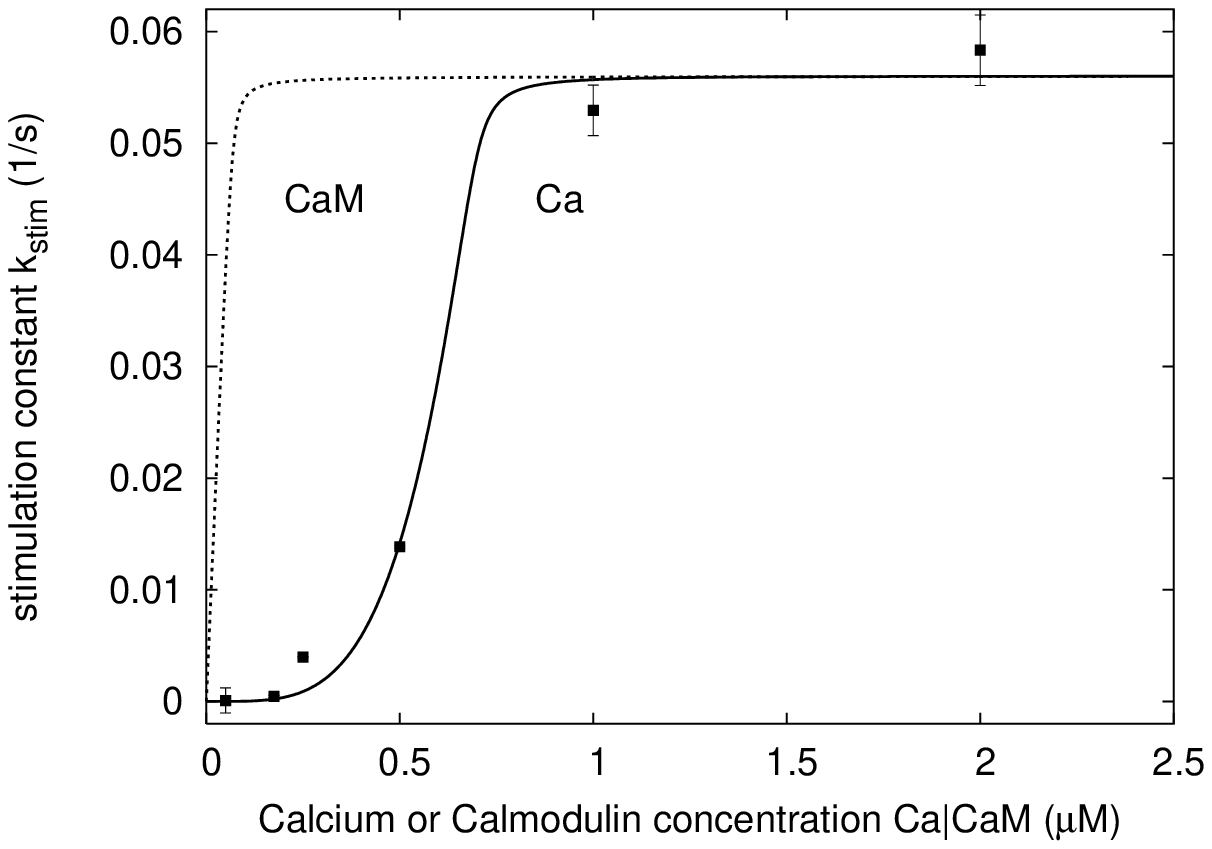, width=0.63\textwidth}}
\subfigure[h4b isoform]{\epsfig{figure=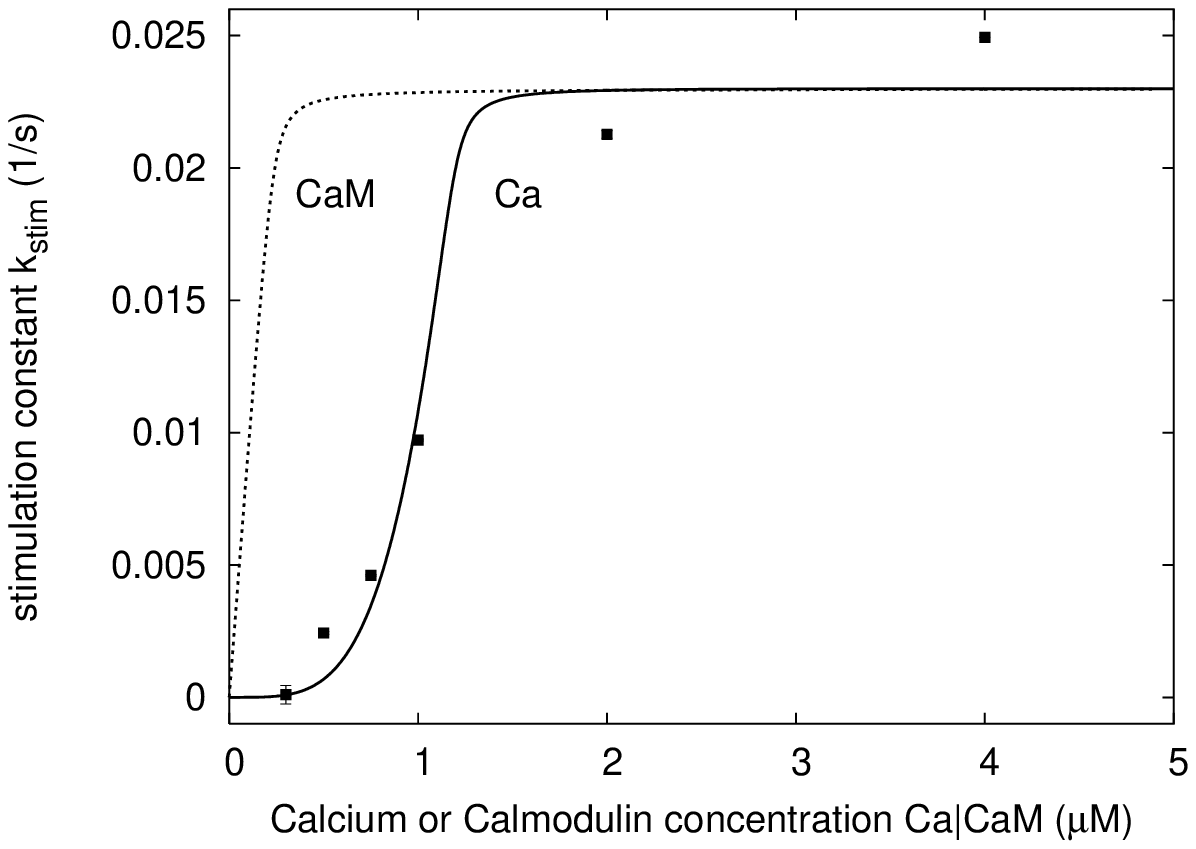, width=0.63\textwidth}} 
\caption{}
\label{fig:k_stim}
\end{figure} 
\begin{figure}[h]
\centering
\subfigure[h2b isoform]{\epsfig{figure=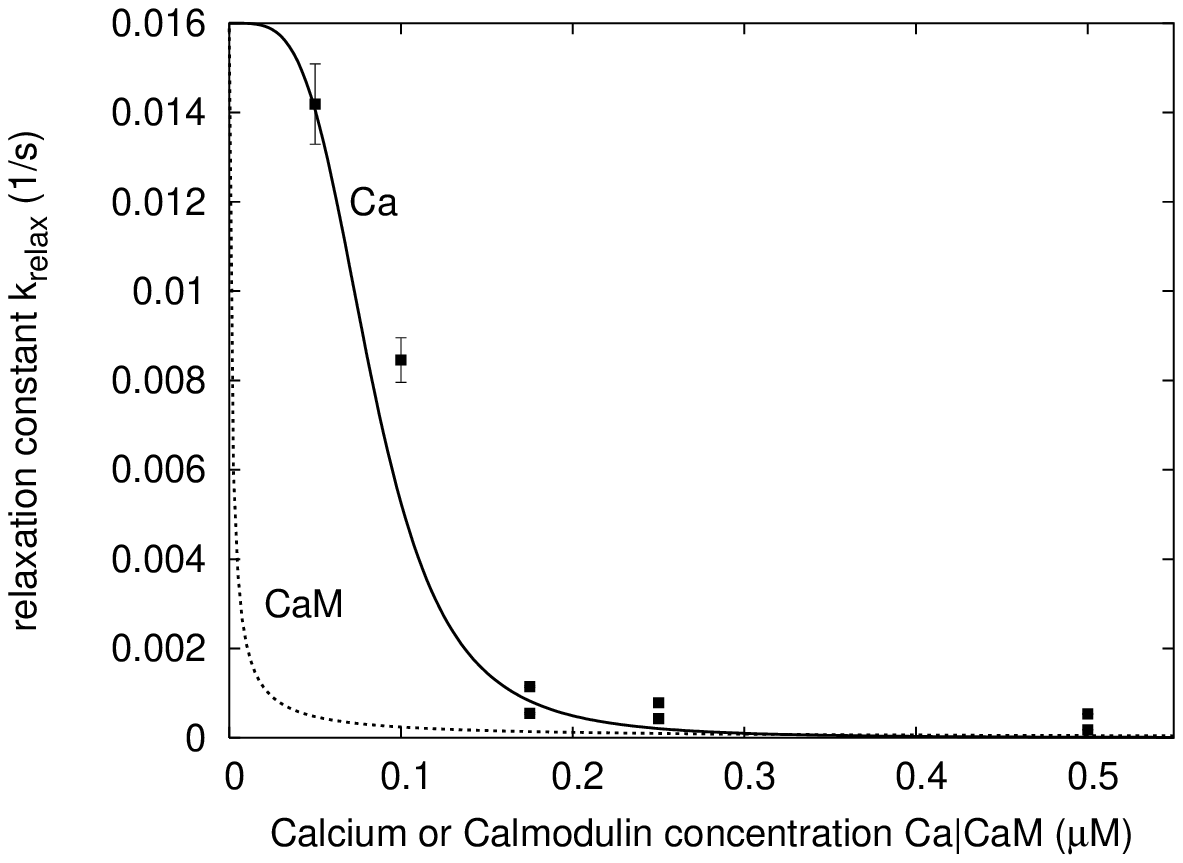, width=0.63\textwidth}}
\subfigure[h4b isoform]{\epsfig{figure=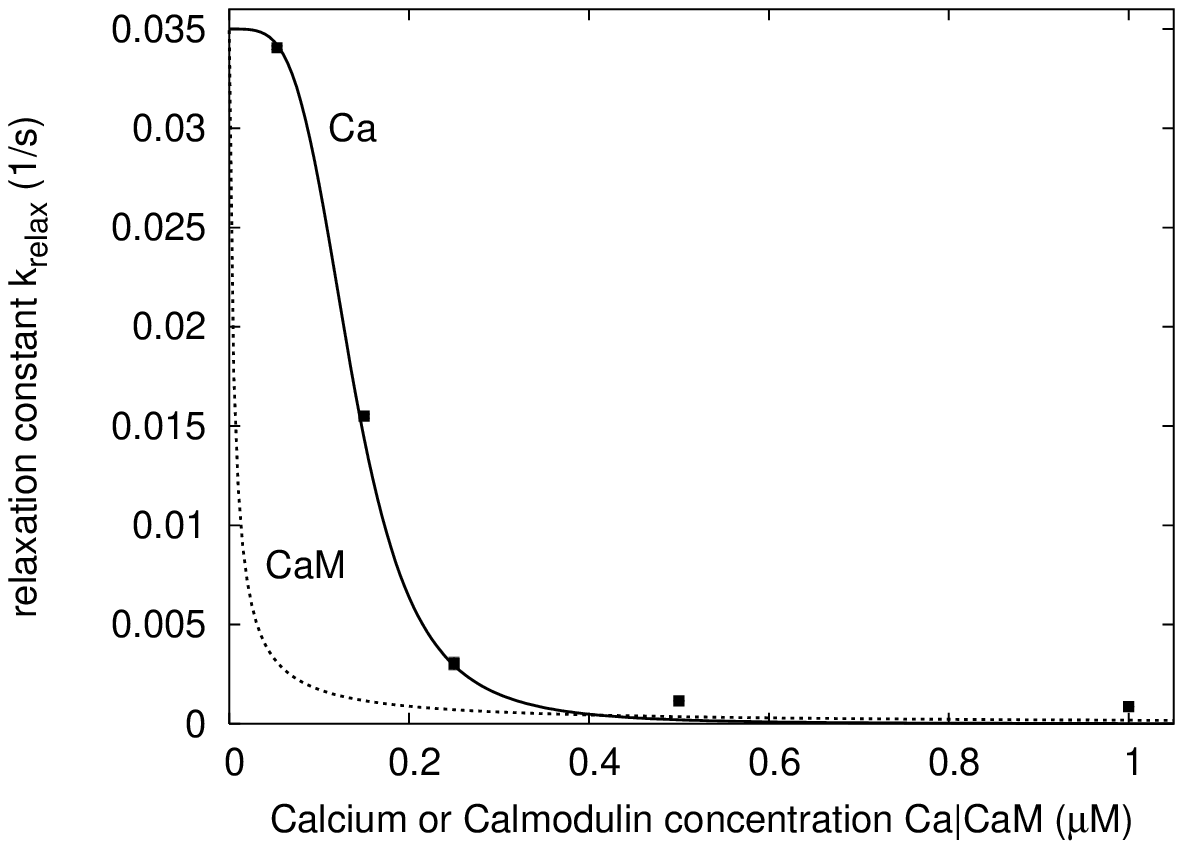, width=0.63\textwidth}}
\caption{}
\label{fig:k_unstim}
\end{figure} 

\begin{figure}[t]%
\centering%
\subfigure[stimulation]{\epsfig{figure=dyn_stim_2b.eps, width=0.55\textwidth}}\\[5mm]
\subfigure[relaxation]{\epsfig{figure=dyn_unstim_2b.eps, width=0.55\textwidth}}
\caption{}
\label{fig:dynamics}%
\end{figure}%
\begin{figure}[t]%
\centering%
\subfigure[h2b isoform]{\epsfig{figure=steady_state_2b_Ca.eps, width=0.55\textwidth}}\\[5mm]
\subfigure[h4b isoform]{\epsfig{figure=steady_state_4b.eps, width=0.55\textwidth}}
\caption{}
\label{fig:steady_state_Ca}%
\end{figure}%
\begin{figure}[t]
\centering
\epsfig{figure=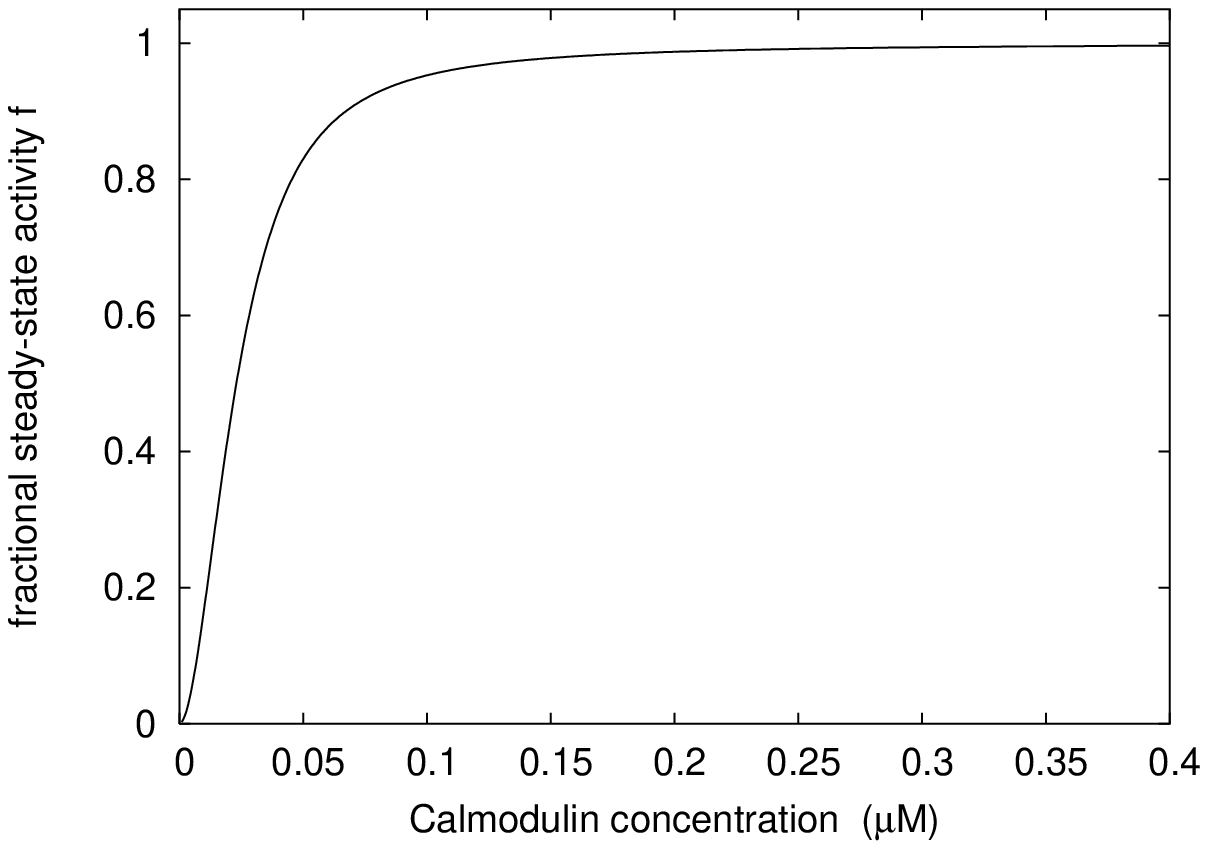, width=0.6\textwidth}
\caption{}
%
\label{fig:steady_state_CaM}%
\end{figure}%
\begin{figure}[t]
\centering
\subfigure[time course of $Ph_i$ production]{\epsfig{figure=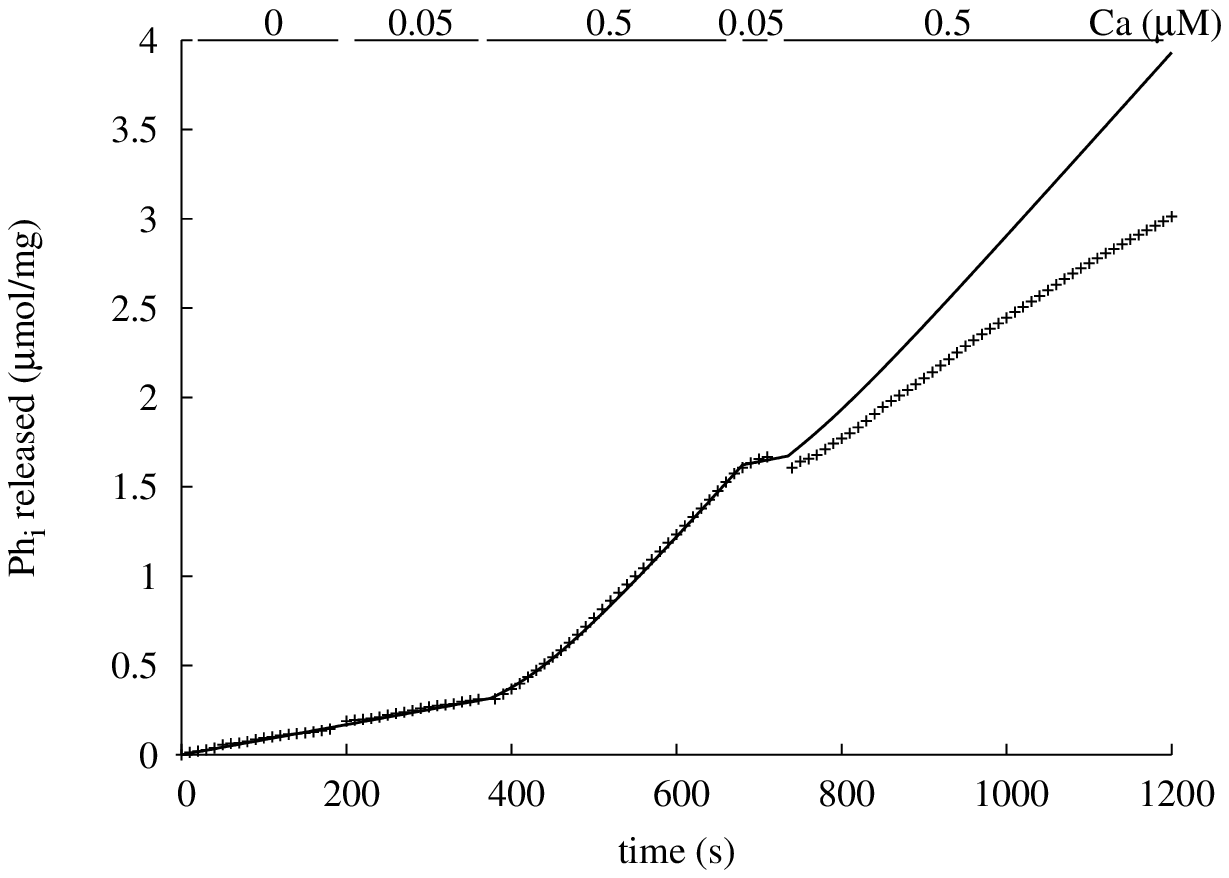, width=0.6\textwidth}}
\subfigure[fraction of change in $J_{0.05}$]{\epsfig{figure=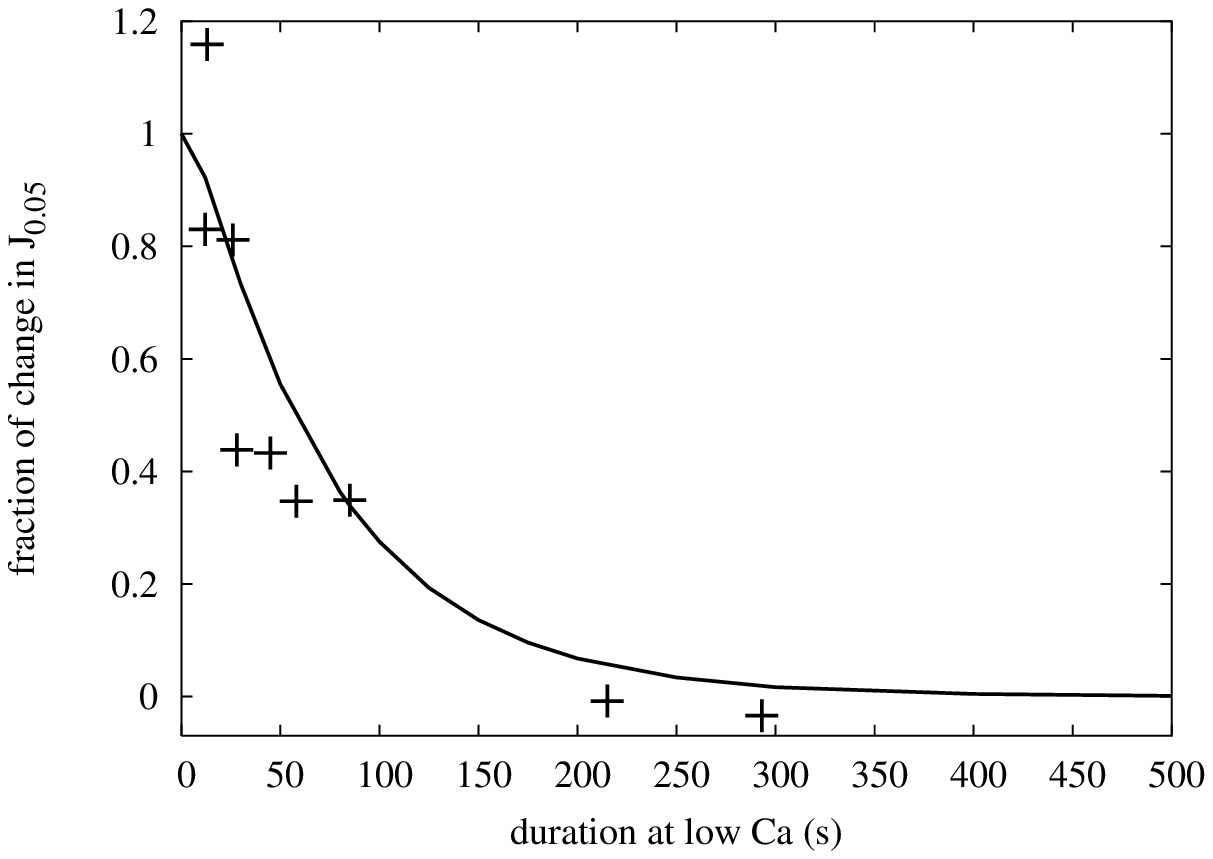, width=0.6\textwidth}}
\caption{}
\label{fig:dyn_activity}
\end{figure}

\end{document}